\documentclass[aps,pre,reprint,superscriptaddress,nofootinbib]{revtex4-1}
\usepackage{graphicx}
\usepackage{amsmath}
\usepackage{amsfonts}
\usepackage{hyperref}
\newcommand{\Wi}{\textrm{Wi}}

\begin{document}

\title{A Force-Level Theory of the Rheology of Entangled Rod and Chain Polymer Liquids. II. Perturbed Reptation, Stress Overshoot, Emergent Convective Constraint Release and Steady State Flow}

\author{Kenenth S. Schweizer}
\email[]{kschweiz@illinois.edu}
\affiliation{Department of Materials Science and Department of Chemistry, University of Illinois, 1304 West Green Street, Urbana, IL 61801, USA}
\author{Daniel M. Sussman}
\affiliation{Department of Physics and Astronomy, University of Pennsylvania, 209 South 33rd Street, Philadelphia, Pennsylvania 19104, USA}

\date{\today}

\begin{abstract}
We numerically and analytically analyze the startup continuous shear rheology of heavily entangled rigid rod polymer fluids based on our self-consistent, force-level theory of anharmonic tube confinement. The approach is simplified by neglecting stress-assisted transverse barrier hopping, and irreversible relaxation proceeds solely via deformation-perturbed reptation. This process is self-consistently coupled to tube dilation and macroscopic rheological response. We predict that with increasing strain the tube strongly dilates, entanglements are lost, and reptation speeds up. As a consequence, a stress overshoot emerges not due to affine over-orientation, but rather to strong weakening of the entanglement network. Just beyond the stress overshoot the longest relaxation time is predicted to scale as the inverse shear rate, corresponding to the emergence of a generic form of convective constraint release (CCR). Its origin is a stress-induced local force that self-consistently couples the mesoscopic tube-scale physics with macroscopic mechanical response. Tube weakening at the overshoot occurs via the same qualitative mechanism that leads to microscopic absolute yielding in the nonlinear elastic scenario analyzed in the preceding paper. The stress overshoot and emergent CCR thus correspond to an elastic-viscous crossover due to deformation-induced disentanglement, which at long times and high shear rates results in a quantitatively sensible flow stress plateau and shear thinning behavior. The predicted behavior for needles is suggested to be relevant to flexible chain melts at low Rouse Weissenberg numbers if contour length equilibration is fast. Quantitative predictions for tube dilation in the steady state flow of chains melts are favorably compared to a recent simulation. The analytic analysis presented in this paper provides new insights concerning the physical origin of our non-classical rheological predictions.
\end{abstract}

\maketitle

\section{Introduction}
In this companion article to the preceding paper I [1], we employ our force-level statistical mechanical approach to study the nonequilibrium and nonlinear \emph{dynamical} effects relevant to the rheology of rigid rod liquids. We focus primarily on finite-rate startup continuous shear deformations in the heavily entangled limit. Of special interest for rods under all flow conditions, and for chains in slow flows where the contour length is equilibrated, is to fundamentally understand the nature of (and possible connections between) the physics of tube dilation, the stress overshoot, emergent convective constraint release (CRR), the steady state flow curve and shear thinning.

As a relevant preliminary, in section II we analyze the deformation-modified terminal relaxation process and tube survival function that quantifies disentanglement in the ``nonlinear elastic'' limit studied in paper I. In section III we implement for rods the full numerical theory for startup continuous shear rheology in a new simplified manner where stress-assisted transverse barrier hopping is not allowed. This simplifies the physical picture underlying our approach, and also allows analytic results (valid for the heavily entangled limit of rod liquids) to be extracted from the coupled constitutive and structural evolution equations. Thus, dissipative relaxation dynamics are \emph{entirely} controlled by a self-consistent treatment of the coupling between reptation, the deformation-modified tube confinement field, and the macroscopic stress-strain response [2]. Key numerical predictions of the theory are established, and physical behavior akin to that proposed in phenomenological CCR models emerges naturally in this treatment. Sections IV and V present analytic derivations of the central numerical results obtained in section III. These allow a greatly enhanced intuitive physical understanding of the origin of our numerical results, the (sometimes surprising) interconnections of different predictions, and the conceptual similarities and differences with the Doi-Edwards (DE)[3] and GLaMM (Graham-Likhtman-McLeish-Milner)[4] models. We propose a partial synthesis of some concepts treated as distinct in phenomenological models. In section VI we briefly discuss how our rheological theory for rods may be relevant to chain polymer liquids if stretch rapidly equilibrates. A numerical application to tube swelling in the steady state, as probed in a recent primitive path (PP) analysis of chain melt simulations [5], is presented in Section VII. The paper concludes in Section VIII with a summary and a look towards the future.

While much of this article stands independently from its companion paper, it has been written assuming the reader is familiar with the preceding paper I [1], and equations from that work are cited as Eq(I.xx). As in paper I, statistical mechanical derivations previously documented in the literature are not repeated.   

\section{Perturbed Reptation and Disentanglement in the Nonlinear Elastic Regime}

Here we bring together our results for entangled rods, and PP chains on timescales both long and short relative to contour-length relaxation, as obtained in paper I to analyze how deformation and orientation modify the reptation-controlled terminal relaxation time and irreversible disentanglement process within the nonlinear elastic scenario [1]. These results are relevant to the early stages of startup shear rheology, and provide context for the remainder of this paper, which goes beyond the nonlinear elastic limit.

\subsection{Rods and Contour-Length Relaxed PP Chains}
As stress or strain grows, tube confinement weakens, entanglements are lost, and thus reptation speeds up, thereby accelerating disentanglement. This change in the disentanglement time scale was not present in the nonlinear elastic limit studied in section III of paper I. Here, we present an ``adiabatic'' analysis of these dynamical questions; in subsequent sections we will show that this analysis is relevant to the full dynamical treatment before the emergence of flow-induced CCR.

In the quiescent state of a heavily entangled rod fluid the terminal rotational time is inversely proportional to the transverse diffusion constant [2,6-8] ${{D}_{\bot }}/{{D}_{\bot ,0}}\propto {{D}_{rot}}/{{D}_{rot,0}}$, and thus:
\begin{equation}
{{\tau }_{rot}}=\frac{{{L}^{2}}}{6{{D}_{\bot }}}=\frac{{{\tau }_{0}}}{36}\cdot \frac{{{D}_{\bot ,0}}}{{{D}_{\bot }}},
\end{equation}
where ${{\tau }_{0}}={{L}^{2}}/{{D}_{||,0}}$ is proportional to the fast (dilute-solution-like for rods) CM longitudinal reptation time. Since here we ignore stress-assisted transverse barrier hopping as an alternate channel for relaxation, one can exploit the same simple connections between ${{\tau }_{rot}}$, ${{D}_{\bot }}$, and ${{d}_{T}}$ that apply in equilibrium and write ${{D}_{rot}}\propto {{D}_{\bot }}\propto \tau _{rot}^{-1}\propto {{({{d}_{T}}/L)}^{2}}$. Using Eqs. (I.7) and (I.20), an approximate representation of the reduction of the relaxation time is thus [2,8]:
\begin{eqnarray}
\frac{{{\tau }_{rot}}(\rho ,S,\sigma )}{{{\tau }_{rot}}(\rho ,0,0)}&=&{{\left( \frac{{{d}_{\bot }}(\rho ,0,0)}{{{d}_{\bot }}(\rho ,S,\sigma )} \right)}^{2}} \nonumber \\
&\approx& {{\left( \sqrt{1-S}-[(\beta \sigma {{L}^{3}})/3(\rho /{{\rho }_{c}})] \right)}^{2}},
\end{eqnarray}
Rod orientation and the direct force effect in the dynamic free energy (transverse tube confinement field) dilate the tube diameter, speeding up relaxation and disentanglement. The approximate equality in Eq. (2) requires that the orientational order parameter ($S$) is not close to unity and the stress ($\sigma$) is not too close to its microscopic absolute yield value [2,8,9], a condition where the neglect of stress-assisted barrier hopping is accurate. 

The above results allow one to analytically derive how the tube survival function changes if the tube diameter slowly changes for any reason:
\begin{equation}
\Psi (\gamma )=\exp \left( -\frac{1}{Wi}\int\limits_{0}^{\gamma }{d\gamma '{{\left( \frac{{{d}_{T}}(S(\gamma ');\sigma (\gamma '))}{{{d}_{T}}} \right)}^{2}}} \right)
\end{equation}
where $\Wi$ is the Weissenberg number. It is the integration through strain history of the reptation form of the relaxation time that we refer to as the ``adiabatic'' scenario. Recall that in equilibrium, or under the DE assumption that there is no change of the relaxation spectrum under deformation, Eq.(3) becomes a simple exponential function
\begin{equation}
{{\Psi }_{DE}}(\gamma )=\exp \left( -\gamma /Wi \right)
\end{equation}
Taking into account only tube weakening due to deformation-induced orientational order (i.e., only the first contribution on the right hand side of Eq.(2), not the term stemming from the ``direct force'' effect), and using Eq.(I.23), we obtain:
\begin{eqnarray}
\Psi (\gamma )&=&\exp \left( -\frac{1}{Wi}\int\limits_{0}^{\gamma }{d\gamma '\left( 1+\left( \gamma {{'}^{2}}/4 \right) \right)} \right) \nonumber \\
&=&\exp \left( -\frac{\gamma }{Wi}\left( 1+\frac{{{\gamma }^{2}}}{12} \right) \right)
\end{eqnarray}
At large strains, disentanglement occurs faster than the exponential form above.

Based on a detailed analysis that includes orientation-induced tube softening and the direct force effect, and using an elastic stress-strain relation $\sigma ={{G}_{e}}\gamma h(\gamma )$, one can derive an approximate result for tube swelling in the nonlinear elastic limit [7]:
\begin{equation}
\frac{{{d}_{T}}}{{{d}_{T}}(\gamma )}\approx 1-\frac{\gamma }{{{\gamma }_{y}}}
\end{equation}
which immediately implies:
\begin{equation}
\frac{{{\tau }_{rot}}(\gamma )}{{{\tau }_{rot,0}}}={{\left( \frac{{{d}_{T,0}}}{{{d}_{T}}(\gamma )} \right)}^{2}}\approx {{\left( 1-\frac{\gamma }{{{\gamma }_{y}}} \right)}^{2}}
\end{equation}
 Using Eqs. (3) and (7) one obtains:
\begin{equation}
\Psi (\gamma )\approx \exp \left( -\frac{{{\gamma }_{y}}}{Wi}\cdot \frac{\gamma }{{{\gamma }_{y}}-\gamma } \right)
\end{equation}
The inverse tube diameter decreases linearly, and the reptation-driven terminal relaxation time decreases quadratically, with strain. As a consequence, an essential singularity enters the tube survival function, corresponding to the abrupt destruction of the transverse confinement field as strain approaches its microscopic absolute yield value.

We expect that these results for rods are relevant for contour-length-relaxed PP chains when reptation is the dominant mechanism of dynamic disentanglement [7,10].

\subsection{Stretched PP Chains}
Implications of unrelaxed chain stretching (i.e., slow contour length retraction after a deformation) and the attendant tube compression for reptation-driven transport were derived in paper I. But what are the strain dependences of the longest relaxation time? Under the assumption of no stretch relaxation, recall the deGennes scaling argument in equilibrium [11]:
\begin{eqnarray}
{{\tau}_{rep}}&\approx &{{\tau }_{0}} \frac{L_{e,contour}^{2}}{{{D}_{Rouse}}} \propto {{\left( \frac{N\sigma }{\sqrt{{{N}_{e}}}} \right)}^{2}}N \nonumber \\
&\propto& \frac{{{N}^{3}}}{{{N}_{e}}}{{\tau }_{0}}\propto {{\tau }_{0}}{{\left( \frac{{{L}_{c}}}{{{d}_{T}}} \right)}^{2}}D_{Rouse}^{-1}
\end{eqnarray}
where the full PP contour length enters in the first term on the right hand side,  but the fixed chemical contour length, $L_c$, enters in the final expression. Equivalently,
\begin{equation}
{{\tau }_{rep}}\propto {{\tau }_{Rouse}}{{\left( \frac{{{R}_{g}}}{{{d}_{T}}} \right)}^{2}}\propto \frac{{{N}^{3}}}{{{N}_{e}}}{{\tau }_{0}}
\end{equation}
How to generalize this to a deformed anisotropic melt at short times is not clear. To proceed, we are guided by the fact that the physical contour length of a real polymer is fixed; what dominates is how much of it must crawl along a coarse-grained tube due to transverse localization. This perspective suggests:
\begin{eqnarray}
{{\tau }_{rep}}(\gamma ) &\propto& {{\tau }_{0}}{{\left( \frac{{{L}_{c}}}{{{d}_{T}}} \right)}^{2}}D_{Rouse}^{-1}\propto \frac{{{N}^{2}}{{b}^{2}}}{d_{T}^{2}(\gamma )}D_{Rouse}^{-1}\nonumber \\
&\propto &{{\tau }_{rep,0}}{{\left( \frac{{{d}_{T}}}{{{d}_{T}}(\gamma )} \right)}^{2}}
\end{eqnarray} 
The final result is of the same form as for rods in Eq. (7): the reptation time changes only via the tube diameter, which if compressed slows down dynamic disentanglement.

Of course, it is unlikely that stretched chains do not relax their contour length on the  reptation time scale. The immediate implications, though, are for the modification of the tube survival function. Using Eqs. (3) and (11) and Eq.(I.47) we obtain:
\begin{align}
  & \Psi (\gamma )=\exp \left( -\frac{1}{Wi}\int\limits_{0}^{\gamma }{dx\frac{{{(1+{{x}^{2}}/4)}^{{}}}}{{{(1+{{x}^{2}}/2)}^{2}}}} \right) \nonumber \\ 
 & \quad \quad \ =\exp \left( -\frac{1}{4Wi}\left( 3\sqrt{2}\arctan (\gamma /\sqrt{2})+\frac{\gamma }{1+{{\gamma }^{2}}/2} \right) \right)
\end{align}
 
At very low strains, Eq.(12) reduces to the exponential from of Eq.(4), as it must. At high strains, the disentanglement rate slows compared to the DE result, and in the very large strain limit dynamical tube destruction is absent. Again, the latter limit is unphysical since Eq. (12) is valid only on time scales short relative to chain stretch relaxation.

\section{Startup Shear Rheology of Entangled Rod Fluids: Dynamical Theory and Numerical Predictions}
\subsection{Formulation}
To fully treat startup continuous shear rheology of rod fluids within our microscopic approach requires taking into account time-dependent tube softening due to both the rod orientation and direct force effects, both of which speed up reptation and viscous disentanglement. A flow-induced disentanglement process emerges from the self-consistent connection between tube scale physics, single-rod motion, and the macroscopic stress-strain response [2]. All these factors are coupled via the spatially-resolved nonequilibrium dynamic free energy (i.e., the tube confinement field) which evolves with strain or time. Given that stress both dilates the tube and lowers the transverse barrier, two channels of polymer motion exist [2,8]: deformation-accelerated reptation and activated transverse barrier hopping. In the heavily entangled limit, we have numerically found that activated hopping is of secondary importance [2], and we ignore it in this work.

The mathematical realization of the above ideas involves coupled evolution equations for stress and rod orientation, and an Òeffective strainÓ concept. We adopt a constitutive equation that is formally of the DE, pure-orientational-stress form [2]:
\begin{equation}
\sigma (t)={{G}_{e}}\int\limits_{-\infty }^{t}{dt'Q(E(t,t'))\frac{d}{dt'}}\Psi (t-t')
\end{equation}
\begin{equation}\tag{13b}
Q(x)\equiv xh(x)\approx 5x/(5+{{x}^{2}})
\end{equation}
\begin{equation}\tag{13c}
\Psi (t-t')=\exp \left( -\int\limits_{t'}^{t}{dt''\frac{1}{{{\tau }_{rot}}(\rho ,S(t''),\sigma (t'')}} \right)
\end{equation}
where the terminal (rotational) relaxation time is given by the first equality in Eq.(2).

Closure of the above equations requires an evolution equation for the orientational order parameter, S(t). We proposed a competition between relaxation-driven orientational randomization and a mechanically-driven, rate-dependent, orientational driving force [2]:
\begin{equation}
\frac{dS(t)}{dt}=\frac{-S(t)}{{{\tau }_{rot}}(\rho ,S(t),\sigma (t))}+\left( {{\left. \frac{d{{S}_{a}}}{d\gamma } \right|}_{\gamma ={{\gamma }_{eff}}}} \right)\dot{\gamma }
\end{equation}
The second term introduces the idea of an effective strain, ${{\gamma }_{eff}}$. Under affine conditions, one has the classic Lodge-Meissner relation: $S(\gamma )=2\gamma {{\left( 3\sqrt{4+{{\gamma }^{2}}}-\gamma  \right)}^{-1}}$ for $S>0$. We postulate that this functional form still applies beyond the affine regime, which provides the self-consistent relation:
\begin{equation}
\gamma_{eff}(t) = \frac{3S(t)}{\sqrt{1+S(t)-2S^2(t)}},
\end{equation}
\begin{equation}
{{S}_{a}}({{\gamma }_{eff}})=S(t).
\end{equation}
where ${{S}_{a}}(\gamma )$ is the rod orientation that results from an affine (step) shear strain deformation of amplitude $\gamma $. This effective strain corresponds to the affine strain needed to generate the current amount of orientational order in the system. The underlying physical idea is that as rod orientational relaxation proceeds, the amount by which further deformation orients rods depends on the current orientational order, not a hypothetical state that exists in the absence of relaxation (e.g., an affine deformed state). In the limit of very high Wi one has:
\begin{equation*}
\tau _{rot}^{-1}\ll \dot{\gamma},\quad {{\gamma }_{eff}}\to \gamma =\dot{\gamma }t
\end{equation*}

Equations (13)-(16), along with the required input from paper I, define the full rheological theory which is numerically solved via integration through deformation history. The tube diameter, relaxation time, orientation parameter, and orientational stress are all coupled via their connection to the nonequilibrium tube confinement field.

\subsection{Numerical Predictions}

As in paper I, we consider a heavily entangled needle fluid with $\rho /{{\rho }_{c}}=1000$, where ${{\rho }_{c}}{{L}^{3}}=3\sqrt{2}$. Figures 1-4 show representative numerical predictions, computed as in our prior work [2] but here neglecting the transverse barrier hopping process.

\begin{figure}
\centerline{\includegraphics[width=0.9\linewidth]{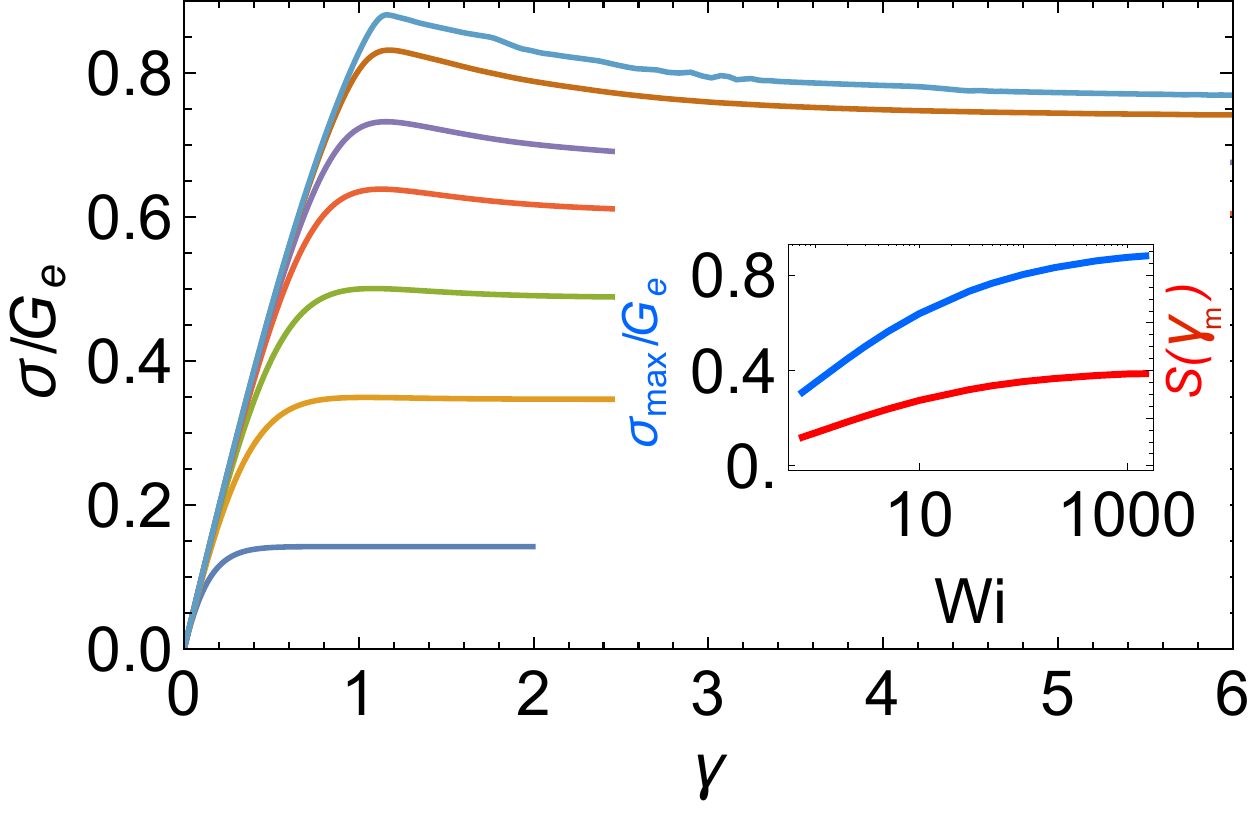}}
\caption{Dimensionless ratio of the shear stress divided by the equilibrium elastic modulus of a heavily entangled needle fluid ($\rho/\rho_c=1000$) as a function of accumulated strain in a startup continuous shear deformation at dimensionless deformation rates of (from bottom to top) Wi $= 0.2$, 1, 3, 10, 30, 200, 1500.  Inset shows the dimensionless stress (upper blue curve) and orientational order parameter (lower red curve) at the stress overshoot as a function of dimensionless shear rate.}
\end{figure}

The main frame of Fig.1 shows the dimensionless stress versus strain at various Wi values. A weak overshoot peak occurs at a strain of order unity, which is very close to its Òabsolute yield strainÓ analog per section III of paper I, and below the DE affine over-orientation value of ${{\gamma }_{m}}\approx \sqrt{5}$. The overshoot amplitude and strain weakly increase with Wi, and the inset shows the corresponding stress and orientational parameter at the overshoot peak. All of these quantities grow monotonically with Wi, and then saturate.

Figure 2 shows results for the tube diameter normalized by its quiescent value versus strain. Massive dilation is predicted, which is largest at the stress overshoot. The magnitude of the dilation, and the tendency for it to exhibit an overshoot, are enhanced as Wi grows. For $\Wi=1000$ and the heavily entangled system studied here, the tube diameter is $\sim 10$ times larger than its equilibrium value.  This is still far smaller than the needle length, and the system remains entangled. For long time transport and relaxation, this implies dynamics $\sim 100$ times faster than in equilibrium.

\begin{figure}
\centerline{\includegraphics[width=0.9\linewidth]{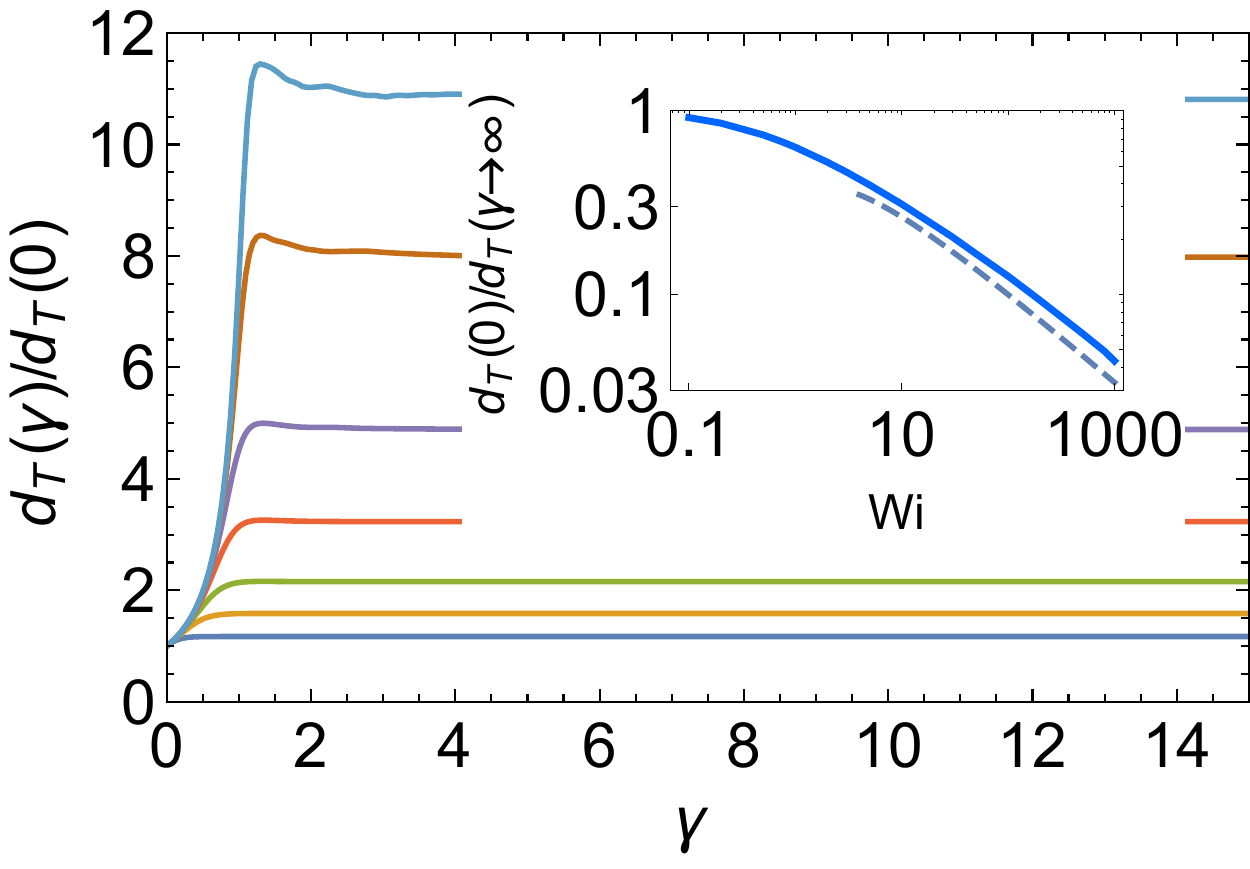}}
\caption{Ratio of the tube diameter at fixed strain to its equilibrium value plotted as a function of strain at various dimensionless shear rates of Wi $= 0.2$, 1, 3, 10, 30, 100, 200 (bottom to top). Inset: Long time steady state value of the same ratio as a function of dimensionless shear rate (solid curve), and the prediction of Eq. (44) (dashed curve).
 }
\end{figure}

Beyond the stress overshoot, the tube diameter attains its steady state value quickly but not instantaneously. The inset shows the ratio of the equilibrium to steady state tube diameter, which decreases strongly as Wi grows due to enhanced tube dilation. Strict power law scaling is not predicted, but over a moderate range of Weissenberg numbers we find $d_T \sim \Wi^{-x}$, $x\approx 0.4-0.5$.

Figure 3 shows the effective terminal relaxation time in units of its equilibrium value as a function of strain at various Wi. The dynamics are greatly enhanced due to tube dilation beginning at strains \emph{well below} the stress overshoot, which occurs at a strain of $\sim 1-1.5$. The functional form agrees well with the parabolic law of Eq.(7) derived under the ``adiabatic'' simplification. Plateau values are attained just beyond the stress overshoot. This signals an emergent CCR regime which is precisely defined here by a relaxation time scaling as the inverse of the shear rate. The inset quantifies this CCR effect as a function of strain in terms of an effective Weissenberg number, $\Wi_{eff} \equiv \dot{\gamma }{{\tau }_{rot}}(\gamma \to \infty )$. At high Wi, we find $\Wi_{eff}\rightarrow 1-2$.

\begin{figure}
\centerline{\includegraphics[width=0.9\linewidth]{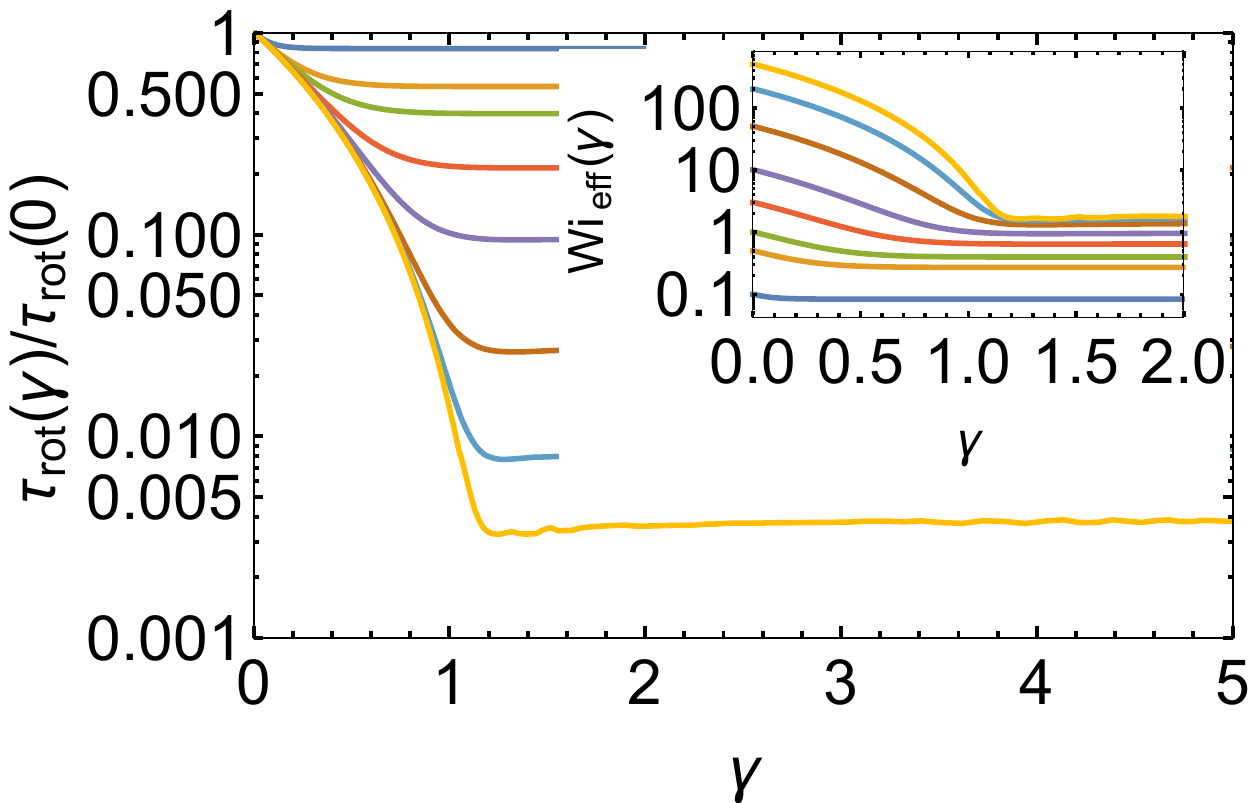}}
\caption{Longest needle relaxation time as a function of accumulated strain, divided by its equilibrium value, for various dimensionless deformation rates (top to bottom) of Wi $=0.1$, 0.5, 1, 3, 10, 50, 200, 500. Inset is the corresponding effective Weissenberg number as a function of strain for the same applied rates as in the main frame (same $Wi$ numbers, ordered from bottom to top).
}
\end{figure}

Figure 4 presents results for the steady state effective Weissenberg number, orientational order parameter, and flow stress as a function of the externally imposed Wi. These quantities follow almost identical dependences. A stable, monotonic flow curve with a near plateau, and hence sensible shear thinning behavior, is predicted. At high Wi values, the steady state flow stress is $\sim 10\%$ smaller, and $S$ is $\sim 25\%$ larger, than at the overshoot.

The overall picture that emerges from the numerical results is that tube dilation, relaxation time acceleration, emergent CCR, and the overshoot feature are \emph{all} intimately related phenomena. Their common physical origin is the self-consistent coupling between macroscopic stress, orientation, and their local force consequences at the anharmonic tube field level. At zeroth order, our new calculations agree well with our prior numerical results [2] for heavily entangled rods that included stress-assisted transverse barrier hopping, thereby establishing the practical unimportance of the latter relaxation channel for heavily entangled systems.
	
\begin{figure}
\centerline{\includegraphics[width=0.9\linewidth]{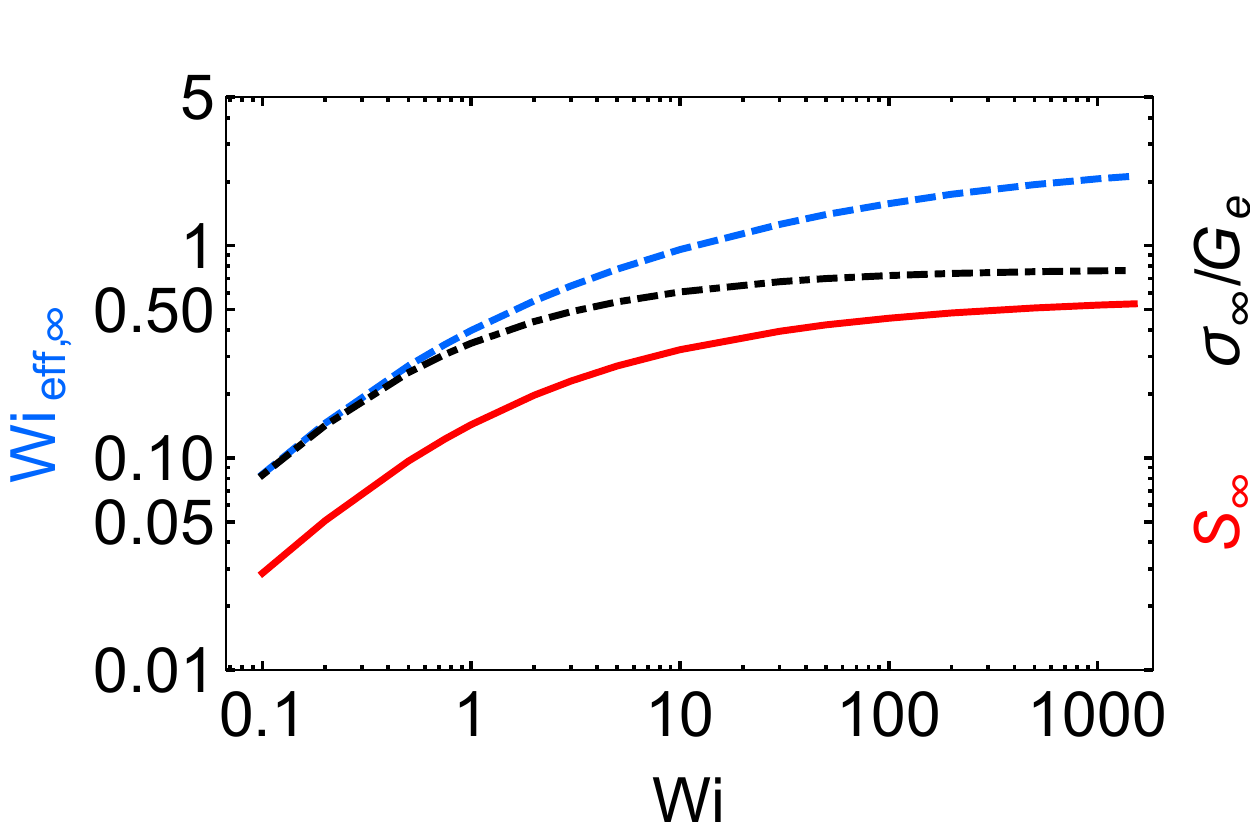}}
\caption{Long time steady state dependence of the effective Weissenberg number (blue dashed), dimensionless flow stress (black dash-dot), and orientational order parameter (red solid) as a function of dimensionless applied shear rate.
}
\end{figure}

None of the trends discussed above agree with DE theory [3], for which the stress overshoot is due to affine over-orientation, the non-monotonic flow curve and excessive shear thinning are unphysical, there is no tube dilation, and the reptation time is invariant to deformation. As a cautionary comment, we emphasize that our suggestion that the stress overshoot is due to emergent CCR, and not affine over-orientation, is sensitive to numerical prefactors of order unity that enter in our approach to nonlinear rheology. We are unaware of attempts to phenomenologically include CCR for entangled rod fluids. Our generic ``emergent CCR'' mechanism Ð a process whereby bulk fluid stress and flow act on the tube scale and modify single polymer dynamics Ð does roughly seem in the qualitative spirit of the phenomenological CCR formulations used for chain rheology. However, our theory qualitatively disagrees with the classic way of incorporating CCR as an additional distinct process which restores a sensible flow curve but is \emph{not} the origin of the stress overshoot. Though quantitative factors matter in our approach, the present results suggest the stress overshoot may be a signature of an elastic-viscous dynamic disentanglement crossover. Even though ``microscopic absolute yielding'' (tube destruction) does not occur here, it is the underlying key concept as manifested by massive tube dilation.

\section{Analytic Analysis of Transient Elastic-Viscous Crossover Regime}
We first analytically analyze the ``pre-flow'' and ``onset of flow'' regimes, which we define as strains up to the stress overshoot. In this regime, tube dilation, growing orientation, faster reptation, and the signatures of emergent CCR all occur.

To make analytic progress, simplifications are invoked. (i) Transverse activated barrier hopping is neglected. (ii) We focus on the heavily entangled regime. (iii) The approximate, but qualitatively reliable, analytic results derived previously and summarized in paper I and above (consistent with point (ii)) are adopted. (iv) The DE-like early time picture of replacing [2,3] continuous startup shear with a step strain form of the constitutive equation is adopted. In the end, the derived results are found to be a faithful representation -- in some cases remarkably so -- of the full numerical calculations for heavily entangled needle fluids presented in section III.

Simplification (iv) is achieved via an integration by parts of Eq.(13). Upon keeping only the strain- (time-) local first term, one obtains:
\begin{equation}
\sigma (\gamma )\approx {{G}_{e}}\frac{\gamma }{1+{{\gamma }^{2}}/5}\Psi (\gamma ;S,\sigma )
\end{equation}
Differentiating to find the maximum yields:
\begin{eqnarray}
\frac{d}{d\gamma }\sigma (\gamma ){{|}_{{{\gamma }_{m}}}}&=&0=\left\{ \frac{1}{1+\gamma _{m}^{2}/5}-\frac{2\gamma _{m}^{2}/5}{{{\left( 1+\gamma _{m}^{2}/5 \right)}^{2}}} \right\} \nonumber \\
& -& \frac{{{\gamma }_{m}}}{1+\gamma _{m}^{2}/5}\cdot \frac{1}{Wi}{{\left( \frac{{{d}_{T}}({{\gamma }_{m}})}{{{d}_{T}}} \right)}^{2}}
\end{eqnarray}
Dropping the last term recovers the classic DE overshoot strain of ${{\gamma }_{m}}=\sqrt{5}$ due to the affine over-orientation effect. Naively, this seems correct if Wi diverges, and it would be \emph{if} the tube diameter did not evolve with strain. However, Eq. (18) is a self-consistent relation for the overshoot strain, and in our approach the tube diameter does swell with increasing deformation. To analyze the full Eq. (18) we first rewrite it as:
\begin{equation}
\frac{{{d}_{T}}({{\gamma }_{m}})}{{{d}_{T}}}=\sqrt{Wi}\sqrt{\frac{5-\gamma _{m}^{2}}{{{\gamma }_{m}}\left( 5+\gamma _{m}^{2} \right)}}
\end{equation}
Since we expect the maximum strain is of order unity, if $\Wi \gg 1$ then Eq. (19) implies a large amount of tube swelling \emph{must} occur, although the tube need not be literally destroyed to achieve flow. As we derive below, the underlying physics controlling the overshoot is deeply connected with emergent CCR.

We now use the result of the second equality in Eq. (2) in Eq. (19) to obtain:
\begin{equation}
\sqrt{1-S({{\gamma }_{m}})}-\frac{1}{3}\frac{{{\rho }_{e}}}{\rho }\tilde{\sigma }({{\gamma }_{m}})=\frac{1}{\sqrt{Wi}}\sqrt{\frac{{{\gamma }_{m}}(5+\gamma _{m}^{2})}{\left( 5-\gamma _{m}^{2} \right)}}
\end{equation}
where $\tilde{\sigma} \equiv \beta L^3\sigma$. If $\Wi \gg 1$ then the right hand side is nearly zero, corresponding to massive tube dilation. As Wi grows, the microscopic absolute yield point is more closely approached, though not reached in this heavily entangled limit. Using the expressions for the orientational order parameter and stress from section III yields a nonlinear equation for $\gamma_m$, and thus $S$. This can be numerically solved, and to leading order in $\Wi^{-1}$ we find:
\begin{equation}
\gamma_m\approx 1,\quad S(\gamma_m)\approx 1/3,\quad \textrm{for}\ \Wi\rightarrow\infty
\end{equation}
These results are consistent with the numerical calculations in Fig. 1, including their saturation at very high Wi. The leading order shift of the overshoot strain with inverse Wi can be straightforwardly computed. We find that this correction is negative and decreases as $\Wi^{-1/2}$. Thus, as $\Wi \gg 1$, the overshoot strain and stress approach their asymptotic limits from below, as found numerically. Eqs. (19)-(21) imply that at the stress overshoot the tube diameter grows as a power law:
\begin{equation}
{{d}_{T}}({{\gamma }_{m}})\ \approx {{d}_{T}}\sqrt{2Wi/3}\propto {{d}_{T}}\sqrt{Wi}
\end{equation} 
The tube is massively swollen, with a form given by the analytic analysis.

The numerical study found CCR-like behavior emerging just beyond the stress overshoot, the origin of which is now clear from the above analysis since
\begin{equation}
\frac{{{\tau }_{rot}}(\sigma ,S)}{{{\tau }_{rot,0}}}={{\left( \frac{{{d}_{T,0}}}{{{d}_{T}}(\sigma ,S)} \right)}^{2}}\approx \frac{2}{3}{{\left( \frac{1}{\sqrt{Wi}} \right)}^{2}}\propto W{{i}^{-1}}
\end{equation}
\begin{equation}
{{\tau }_{rot}}({{\gamma }_{m}})\approx \frac{2}{3Wi}{{\tau }_{rot,0}}=\frac{2}{3}{{\dot{\gamma }}^{-1}}
\end{equation}
Thus, the overshoot and emergence of CCR occur at essentially the same stress or strain, and are intimately related via the massive tube dilation driven by its coupling to the macroscopic stress via the direct force in the dynamic free energy of Eq(I.13). Equation (23) shows that CCR effectively emerges smoothly from the reptation time as the tube diameter becomes flow rate dependent. These features conflict with the DE scenario [3]. They are also unlike the GLaMM model [4] for chain melts which independently accounts for the overshoot via the affine over-orientation effect and the plateau flow stress via empirical ``fine tuning'' [15] of the parameter which quantifies the CCR mechanism: $c_\nu$, the number of retraction events required to result in a tube hop of order the tube diameter [7]. 
 	
In the pre-overshoot regime, one can derive the dependence of the reptation time on accumulated strain. Combining the above results, we find to a good approximation that the reptation time decreases as:
\begin{equation}
\frac{{{\tau }_{rot}}(\gamma )}{{{\tau }_{rot,0}}}\approx {{\left( 1-\frac{\gamma }{{{\gamma }_{m}}} \right)}^{2}}, \quad\gamma <{{\gamma }_{m}}
\end{equation}
This ``parabolic law'' agrees well with our numerical results and the ``adiabatic'' analysis in section II that led to Eq.(7). An empirical interpolation of Eqs. (24) and (25) is:
\begin{equation}
\frac{{{\tau }_{rot}}(\gamma )}{{{\tau }_{rot,0}}}\approx {{\left( 1-\frac{\gamma }{{{\gamma }_{m}}} \right)}^{2}}+\frac{2}{3Wi}\quad \quad 
\end{equation}
This implies an effective terminal relaxation time of:
\begin{equation}
{{\tau }_{rot}}\approx {{\tau }_{rot,0}}{{\left( 1-\frac{\gamma }{{{\gamma }_{m}}} \right)}^{2}}\ +\ \frac{2}{3\dot{\gamma }}
\end{equation}
The apparent additive form mirrors Marrucci's simple CCR formula [13]. Here, however, the flow rate term is \emph{not} added as an independent convective process that ``sweeps away'' tube constraints. Rather, reptative motion continuously accelerates due to tube softening. In essence, CCR emerges continuously as the self-consistent ``end point'' of tube softening due to the nonlinear feedback between the anharmonic confinement field, polymer motion, and macroscopic mechanical response. 

The above analysis also establishes that the simplest form of the CCR-controlled tube survival function, $\Psi (\gamma )\approx \exp \left( -3\gamma /2 \right)$, applies in the vicinity of the stress overshoot, but \emph{not} below it. This has two important implications: (i) well below the stress overshoot is a \emph{nonlinear-elastic-like} regime since CCR is not fully operable, and (ii) the stress overshoot is an \emph{elastic-viscous crossover}. These points seem in qualitative contrast with the DE [3] and GLaMM [4] approaches for chain melts where CCR is ``turned on'' at early times and plays no role in determining the stress overshoot at low $\Wi_R$.

Finally, our full dynamical analysis predicts an overshoot strain very close to the simple absolute yield estimates in section III. This is not an accident, since in the former case we predict that the overshoot (onset of stress drop) is correlated with emergent CCR and a near destruction of the tube (i.e., massive tube dilation and entanglement reduction). This is an important point Ð even in dynamic startup shear, the concept of a ``microscopic absolute yielding'' of a nonlinear elastic origin is relevant.   

\section{Analytic Analysis of the Long Time Steady State Flow Regime}
We now analytically consider the inter-relationships between the flow curve and the steady state values of the rod orientation, tube diameter, and relaxation time. From Eq. (13), and the existence of a time-independent steady state, one has the self-consistent nonlinear relation:
\begin{equation}
{{\sigma }_{\infty }}={{G}_{e}}Q({{\gamma }_{eff,\infty }})\dot{\gamma }{{\tau }_{eff,\infty }}={{G}_{e}}Q({{\gamma }_{eff,\infty }})\ \Wi_{eff,\infty }
\end{equation}
where the effective steady state Weissenberg number depends on the steady state stress and orientational order parameter. The effect of orientation on this stress enters via the damping function evaluated using our effective strain concept [2]. To proceed, we take the steady state limit of the evolution Eq. (14):
\begin{equation}
\frac{dS}{dt}=-\frac{1}{{{\tau }_{eff}}}S+\dot{\gamma }{{\left( \frac{\partial {{S}_{a}}}{\partial \gamma } \right)}_{\gamma \equiv {{\gamma }_{eff}}}}=0
\end{equation}
which implies
\begin{equation}
\Wi_{eff,\infty }={{S}_{\infty }}{{\left( \frac{\partial {{S}_{\infty }}}{\partial {{\gamma }_{eff}}} \right)}^{-1}}
\end{equation}
\begin{equation}
{{S}_{\infty }}=S({{\gamma }_{eff,\infty }})=\frac{2{{\gamma }_{eff,\infty }}}{{{\gamma }_{eff,\infty }}-3\sqrt{4+\gamma _{eff,\infty }^{2}}}
\end{equation}
Using Eq. (2) and the above relations in Eq. (28) one obtains:
\begin{equation}
\frac{{{\sigma }_{\infty }}}{{{G}_{e}}}=Q({{S}_{\infty }})\ \Wi\ {{\left( \sqrt{1-{{S}_{\infty }}}-\frac{3\sqrt{2}}{5}\frac{{{\sigma }_{\infty }}}{{{G}_{e}}} \right)}^{2}}
\end{equation}
This nonlinear equation for the dimensionless flow stress includes all the tube softening effects. Equation (32) is not closed since it requires the steady state polymer orientation, which depends on the steady state stress. But one can first formally solve the quadratic equation for the dimensionless flow stress to obtain:
\begin{eqnarray}
\frac{{{\sigma }_{\infty }}}{{{G}_{e}}}=\frac{25}{36}\Bigg{[}&&\frac{1}{{{Q}_{\infty }}Wi}+\frac{6\sqrt{2}}{5}\sqrt{1-{{S}_{\infty }}} \\
&\pm &\left. \sqrt{{{\left( \frac{1}{{{Q}_{\infty }}Wi}+\frac{6\sqrt{2}}{5}\sqrt{1-{{S}_{\infty }}} \right)}^{2}}-\frac{72}{25}(1-{{S}_{\infty }})} \right] \nonumber
\end{eqnarray}
Taking the negative root, and simplifying in the $W \gg 1$ limit, gives:
\begin{equation}
\frac{{{\sigma }_{\infty }}}{{{G}_{e}}}=\frac{5\sqrt{2}}{6}\sqrt{1-{{S}_{\infty }}}-\frac{25}{36}\sqrt{\frac{12\sqrt{2(1-{{S}_{\infty }})}}{5{{Q}_{\infty }}\Wi}}+.....
\end{equation}
One sees there is an intrinsic limit as Wi diverges (flow stress plateau, perfect shear thinning) which is approached from below, as we found numerically. Substituting Eq.(34) into Eq.(29), the leading order contributions cancel and one obtains:
\begin{eqnarray}
\Wi_{eff}&=&\Wi\ {{\left( \sqrt{1-{{S}_{\infty }}}-\frac{3\sqrt{2}}{5}\frac{{{\sigma }_{\infty }}}{{{G}_{e}}} \right)}^{2}} \\
&=&Wi{{\left( \frac{25}{36}\sqrt{\frac{12\sqrt{2(1-{{S}_{\infty }})}}{5{{Q}_{\infty }}Wi}} \right)}^{2}}=\frac{125\sqrt{2}}{108{{Q}_{\infty }}}\sqrt{1-{{S}_{\infty }}}\nonumber
\end{eqnarray}
Thus, we deduce that the effective steady state Wi number is a rate-independent number. This implies the emergent CCR-like prediction:
\begin{equation}
{{\tau }_{eff}}={{\dot{\gamma }}^{-1}}\left( \frac{125\sqrt{2}}{108{{Q}_{\infty }}}\sqrt{1-{{S}_{\infty }}} \right)
\end{equation}

We note that Eqs. (35) and (36) express the steady state stress and CCR relaxation time in terms of steady state orientation. An implication is that the prefactor of the emergent CCR rate can be understood explicitly as:
\begin{eqnarray}
{{\tau }_{eff}}&=&{{\dot{\gamma }}^{-1}}\left( \frac{125\sqrt{2}}{108{{Q}_{\infty }}}\cdot \frac{6}{5\sqrt{2}}\cdot \frac{{{\sigma }_{\infty }}}{{{G}_{e}}} \right) \\
&=&{{\dot{\gamma }}^{-1}}\left( \frac{25}{18}\cdot \frac{{{\sigma }_{\infty }}}{{{G}_{e}}{{Q}_{\infty }}} \right)=\frac{25}{18}{{\dot{\gamma }}^{-1}}\left( \frac{{{\sigma }_{\infty }}}{{{G}_{e}}{{\gamma }_{eff,\infty }}}h({{\gamma }_{eff,\infty }}) \right) \nonumber
\end{eqnarray}
Thus, besides the numerical prefactor, the amplitude of the CCR time is related to what can be interpreted as the ratio of the flow stress to an effective elastic stress including the damping function, evaluated at the ``effective steady state strain.'' It appears that the quantitative aspects of emergent CCR is a rather intricate issue that is functionally coupled to flow stress, the bare elastic modulus, and effective strain/over-orientation. These insights may be germane to the uncertainty concerning how to precisely quantify the CCR effect in phenomenological models [12,15].

 The prefactor in Eq. (36) and the flow stress can be explicitly determined from the steady state orientational order parameter, or equivalently from the effective strain. This calculation is done using the expression for $\Wi_{eff}$ in Eq. (35), taking the derivative of $S$ with respect to the effective strain, and then simplifying. The required algebraic manipulations employ the following readily-derived relations:
\begin{equation}
\Wi_{eff} =\frac{125\sqrt{2}}{108}\cdot \frac{\sqrt{1-{{S}_{\infty }}}}{{{Q}_{\infty }}}\quad \equiv \quad {{S}_{\infty }}\left( \frac{\partial {{\gamma }_{eff,\infty }}}{\partial {{S}_{\infty }}} \right)
\end{equation}
\begin{eqnarray}
{{\gamma }_{eff,\infty }}&=&\frac{3{{S}_{\infty }}}{\sqrt{1+{{S}_{\infty }}-2S_{\infty }^{2}}} \\
& \Rightarrow & \left( \frac{\partial {{\gamma }_{eff}}}{\partial {{S}_{\infty }}} \right)=\frac{3\left( 1+\frac{{{S}_{\infty }}}{2} \right)}{{{\left( 1+{{S}_{\infty }}-2S_{\infty }^{2} \right)}^{3/2}}} \nonumber
\end{eqnarray}
\begin{equation}
{{Q}_{\infty }}=\frac{{{\gamma }_{eff,\infty }}}{1+\frac{1}{5}{{({{\gamma }_{eff,\infty }})}^{2}}}=\frac{3{{S}_{\infty }}}{\left( 1+{{S}_{\infty }}-\frac{1}{5}S_{\infty }^{2} \right)\sqrt{1+{{S}_{\infty }}-2S_{\infty }^{2}}}
\end{equation}
Combining the above results and simplifying, one obtains a closed equation for the steady state orientational order parameter:
\begin{eqnarray}
\frac{125\sqrt{2}}{108}\sqrt{1-{{S}_{\infty }}}&&\left( 1+{{S}_{\infty }}-\frac{1}{5}S_{\infty }^{2} \right) \\
&=&9S_{\infty }^{2}\left( 1+\frac{{{S}_{\infty }}}{2} \right){{\left[ 1+{{S}_{\infty }}-2S_{\infty }^{2} \right]}^{-1}} \nonumber
\end{eqnarray}
Solving this numerically yields:
\begin{equation}
{{S}_{\infty }}\approx 0.41\quad >\ \quad S({{\gamma }_{m}})\approx 0.33
\end{equation}
The deduced value is sensible compared to our numerical calculations, including a value of $S$ modestly larger ($\sim 25\%$) in steady state compared to at the overshoot. 

Using the result in Eq.(22) in the above analytic relations yields quantitative  predictions for steady state properties in the large Wi limit. The results are:
\begin{align}
& {{\gamma }_{\infty }}\approx 1.20\quad ,\quad {{Q}_{\infty }}\equiv {{\gamma }_{\infty }}h({{\gamma }_{\infty }})\approx 0.78 \\ 
 & \frac{{{\sigma }_{\infty }}}{{{G}_{e}}}\approx 0.9\ \quad ,\quad \frac{{{d}_{T,\infty }}}{{{d}_{T}}}\approx 0.94\sqrt{Wi}\nonumber \\ 
 & W{{i}_{eff,\infty }}\approx 1.6\quad ,\quad {{\tau }_{eff,\infty }}\approx \frac{1.6}{{\dot{\gamma }}} \nonumber
\end{align}
The effective strain and $Q$-function are modestly larger than their values at the stress overshoot. The flow stress in units of the equilibrium shear modulus, and the effective Wi number, are sensibly of order unity, and modestly less than at the stress overshoot. The effective relaxation time scales as the inverse of the flow rate. The tube diameter scales as $\Wi^{1/2}$, qualitatively identical to its behavior at the stress overshoot. 

The above analytically derived results all agree well with the numerical calculations in section III, and thus provide a mathematically and physically clear picture of their origin. One difference might appear to be that the analytically obtained power law scaling of the steady state tube diameter does not fully agree with the numerical results in the inset of Fig.2. This is a consequence of the slow approach to the asymptotic large Wi limit, which can be understood by including the leading order correction to the tube diameter scaling in Eq. (43). Using the second equality in Eq. (2) and Eqs. (32)-(34), one obtains:
\begin{equation}
\frac{{{d}_{T,\infty }}}{{{d}_{T}}}=\frac{1.07}{\sqrt{\Wi}}-\frac{0.75}{\Wi}+....
\end{equation}
This result captures the pre-asymptotic behavior as shown in the inset of Fig.2.

Finally, we consider the tube survival function in the steady state. Equations (43) and (26) yield:
\begin{equation}
\Psi (\gamma )\approx \exp \left( -b\cdot \gamma  \right)\ \quad ,\ \gamma >>{{\gamma }_{m}}
\end{equation}
where $b\sim 1$. This form is employed in the simplest versions of CCR in phenomenological models [4,13]. Again the steady state behavior differs little from what is predicted just beyond overshoot since the overshoot is weak and CCR sharply emerges there.

Conceptually, we re-emphasize that in our approach, CCR emerges in a predictive manner beyond the characteristic time (or strain) scale associated with the stress overshoot as a consequence of our self-consistent single-polymer dynamic theory for the anharmonic tube confinement field and a rather simple rheological constitutive formulation. The intuitive heart of our calculation rests on the idea that as polymers move transversely the tube constraints nonlinearly soften, and vice-versa [2]. What we call the ``direct force'' contribution to the dynamic free energy that controls transverse polymer displacements is related to the macroscopic stress. Thus, the tube confinement field controls how polymers move, but the motion of polymers determine how the tube confinement field evolves. This appears to be a conceptually different physical picture than embedded in the GLaMM [4] and other phenomenological models. Furthermore, our emergent CCR mechanism cannot be related to chain retraction since the rods are of fixed length. Instead, we have formulated a theory in which deformation, shear rate, and flow modify the dynamic confinement field implemented in a single particle microrheological spirit. This idea is general in that it transcends material type (molecules, atoms, colloid, polymers) and the physical origin of the dynamical confinement (caging in glasses, physical bonds in gels, entanglements, etc.) [2,8,9,16-18]. Of course, for flexible coil liquids, chain retraction may indeed trigger CCR, and the quantitative strength of tube-softening-based CCR versus chain-retraction-based CCR for these systems is an open issue in our approach.

\section{Startup Continuous Shear Rheology of Chain Melts }
What is the connection between our analysis of rod shear rheology and that of liquids composed of contour length relaxed d-PP chains? Qualitatively, we expect all results carry over within the IAA simplification [3] assuming (as in paper I) a rubber-like network picture in which the entanglement elastic modulus is not perturbed by strain [1]. If this is true, then the microscopic absolute yield results of paper I also carry over, and we obtain numerically sensible predictions for the nearly Wi-independent location of the stress overshoot which are close to its microscopic absolute yield estimate. Similarly, emergent CCR, and its implications on the flow curve and shear thinning, carry over. Importantly, the rod-based results for these quantities are numerically sensible for chain melts in slow flows [12,15], e.g., ${{\sigma }_{\infty }}\approx 0.9{{G}_{e}}$. Of course, at a fully quantitative level, differences between rods and the d-PP chain model must be present, the analysis of which is beyond the scope of this paper. For the general case where contour length equilibration in chain liquids is not assumed, the problem is more complicated since an evolution equation for chain stretch is required and both stretch and orientational degrees of freedom contribute to the stress. This is a direction for future work, but to the extent that chains recover their equilibrium contour length quickly in slow flows, one might expect some or many of the results presented in this article for rods to be qualitatively applicable.

\section{Tube Dilation in Steady State Flowing Entangled Chain Melts}
The above considerations of the relation of the nonlinear rheology of rods and flexible chains motivates an attempt to quantitatively compare our theory with the recent simulations [5] for the steady state behavior of atomistic models of polyethylene melts of moderate degree of entanglement $Z\sim 14$. Baig et.al. separately determined how both the effective entanglement density and the degree of orientational order varied over a wide range of Wi in the flow state [5]. As noted above, the effect of stress is likely more subtle in chain systems compared to rods, and so here we explore the simplest idea that in the steady state the relation between tube diameter and $S$ that we derived in equilibrium still applies. We neglect, however, the direct stress effect on the tube diameter. For our rigid rod theory this is a radical simplification, and here it reflects the fact that our model does not uniquely specify how to translate macroscopic stresses to the PP scale for flexible chains. Clarifying this issue is a future goal.

	In Eq. (I.29) we noted that the tube diameter in the d-PP model, considering only orientational ordering of the primitive paths and neglecting any contributions from stress, can be written as
\begin{equation}
1=\frac{{{\rho }_{PP}}L_{e}^{3}}{16{{\pi }^{2}}\sqrt{2}}\,F\left( \frac{{{L}_{e}}}{{{r}_{l}}} \right)\,G
\end{equation}
where the functions $G$ and $F(x)$ are discussed in paper I. Thus, given the joint probability distribution for the orientation of the PP segments (which determines $G$), one can explicitly compute the orientation-induced softening of the tube diameter. 

Since the simulation study [5] only reported a scalar measure of the orientational distribution of PP steps -- the order parameter $S$ -- and not the full distribution, we employ two simple estimates of the orientational distribution. As a first estimate, we assume that the polymer orientational distribution, $\alpha ({{\vec{u}}_{j}})$, is that of a nematic fluid with a given value of $S$ [9]. As a second approach, we assume orientation is induced by an affine step shear deformation of amplitude $\gamma $, and take the associated degree of order to be $S=-2\gamma /(\gamma -3\sqrt{{{\gamma }^{2}}+4})$ [9]. This captures the intuition that the orientational effects in steady state shear are likely better captured by a distribution with the same symmetry.

\begin{figure}
\centerline{\includegraphics[width=0.9\linewidth]{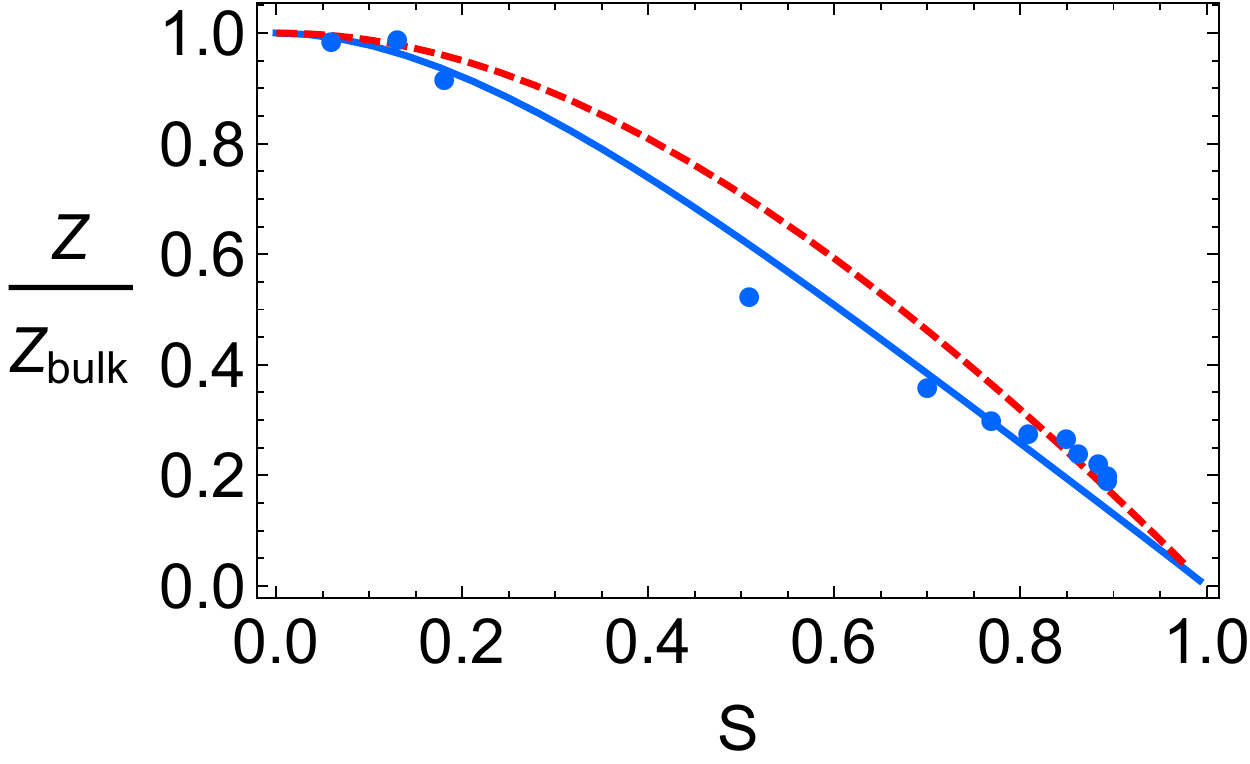}}
\caption{Effective degree of entanglement as a function of orientational order parameter in the steady state. Points are the simulation results of Baig et al.[5]. Curves are the theoretical calculations described in the text, where the solid blue curve corresponds to the step-shear case and the dashed red curve to the Onsager-like nematic order parameter case.
}
\end{figure}
 
We plot the results of these calculations in Fig. 5 in terms of the relative reduction of entanglement density, $Z/{{Z}_{bulk}}={{({{d}_{T}}(0)/{{d}_{T}}(S))}^{2}}$, together with the simulation data. We find that both theoretical curves (with no adjustable parameters) are in reasonably good agreement with simulation. We also find that the curves are only modestly sensitive to the details of the orientational distribution used to determine the value of $S$. Since the correct steady state orientational distribution is almost certainly different from both of our simple models, we take this as an encouraging sign that given our theory would also produce reasonable results if we had used the exact orientational distribution. We note that a recent modification of the CCR idea within the phenomenological tube model framework [14], different than our work, also appears to be consistent with these simulation results [5].

\section{Summary and Discussion }

 	Our primary focus in this article has been the numerical and analytic analysis of startup continuous shear rheology based on our force-level, self-consistent, anharmonic tube theory of entangled rod polymers. Although the entropic barrier to transverse motion, and hence the tube confinement field, can be destroyed at a critical stress in our approach (``absolute microscopic yield''), this effect was neglected in order to focus on our theoretical predictions based on longitudinal reptation (perturbed by, and coupled to, macroscopic deformation and orientation) as the sole origin of terminal relaxation and dissipation. The neglect of stress-assisted transverse barrier hopping, which makes the derivation of analytic results possible, is appropriate for heavily entangled needle fluids in regimes where the fast flows do not generate enough stress to strongly decrease the dynamic free energy barrier to transverse motion. 

For rigid needles, the theory predicts that with increasing strain the tube dilates, entanglements are lost, and reptation speeds up. We further predict that the stress overshoot has its origin in the massive weakening of the entanglement network, not in the affine over-orientation effect as in traditional tube model approaches. We note that this non-classical conclusion is sensitive to numerical prefactors of order unity that enter our approach. Beyond the overshoot, a form of CCR robustly emerges, corresponding to a scaling of the terminal relaxation time as the inverse shear rate. Its physical origin is mechanically-accelerated reptation via a degree of tube dilation that is close to its steady state value. The overshoot location is only weakly rate-dependent, and although the tube is not destroyed for the heavily entangled systems analyzed here, it is massively weakened via the same mechanism underlying microscopic absolute yielding in the nonlinear elastic scenario of paper I. 

Thus, we are lead to a qualitatively new view of the possible origin and meaning of the stress overshoot compared to existing phenomenological tube models: it arises due to strong tube dilation and is an elastic-viscous crossover due to deformation-induced disentanglement. Precisely this same mechanism leads to our prediction of a stable flow curve, including a quantitatively sensible value of the flow stress plateau and degree of shear thinning.  While we have contrasted our approach with the physics of existing phenomenological CCR models, convective-constraint-release ideas have not been developed for rod systems as they lack the chain retraction mechanism underlying CCR in flexible coil melts. We expect that our mechanism of generating emergent CCR-like physics Ð as a consequence of external stress inducing a local force on the tube which couples the mesoscopic scale tube physics with the macroscopic rheological response Ð is a generic one that is neither limited to rigid rod systems nor necessarily tied to chain retraction nor other specific features of chain melts. Indeed, the same basic mechanism accounts well for a (near) stress plateau and shear thinning in colloidal glasses [16] and gels [18], and polymer glasses [17], within the nonlinear Langevin equation theory framework as applied to systems where relaxation and flow is controlled by stress-assisted thermally activated dynamics. 

The predicted behavior for needles is argued to qualitatively apply to flexible chain melts in the low Rouse Wi regime where a PP description built on rapid contour length equilibration is plausible.  In this slow flow regime, our theoretical results seem qualitatively consistent with the non-classical ideas of Wang and coworkers of deformation-induced disentanglement [19-21]. Our work also provides a theoretical basis for their speculative arguments [19-21] concerning the finite ``cohesion'' of the tube, here in the precise and restricted sense of \emph{transverse} entanglement localization.

 	Our predictions for the connections between tube dilation and the rod orientational order parameter under affine deformation conditions (nonlinear elastic scenario [1]) were shown to be in reasonable accord with steady state simulations of an entangled chain polymer melt. This connection is in the spirit of our finding that a full dynamical treatment of the stress-strain response exhibits features below and near the stress overshoot which are well anticipated by the nonlinear elastic limit analysis of paper I based on global affine deformation and a deformed anharmonic tube field. 
   
	From a broad perspective, we believe that our force-level, self-consistent approach addresses the following three questions (in the context of rods and contour-length-relaxed PP chains) posed by McLeish [12] regarding key foundational issues for entangled polymer linear and nonlinear rheology. (i) What is the nature of the tube confinement field? (ii) Can one relax the assumption that the tube confining field is unchanged in nonlinear flow? (iii) What is the correct nonlinear physics of constraint release in strong flows? 

	Future experimental tests of our theory for rod rheology in startup shear should focus on heavily entangled stiff polymer systems such as F-actin, or even more rigid systems like microtubules or other well-dispersed model rigid rod fluids. Novel active microrheology studies of entangled F-actin solutions [22,23] have recently appeared with many fascinating observations reported under interrupted startup shear conditions which contain aspects of nonlinear step strain and continuous shear bulk rheology measurements. The experimental data has been interpreted as showing multiple non-classical tube model features, which are suggested to be qualitatively consistent with our theory. A partial list includes: entanglement network yielding at strains of order unity, tube tightening at early times followed by dilation, unusual power law kinetics of entanglement network healing after yielding, and shear thinning. Highly non-classical behavior was also reported by the same group for entangled DNA solutions [24]. However, we caution that F-actin is not a rigid rod, and the mechanism for storing stress and transverse tube localization involves physical aspects associated with polymer semiflexibility that differ from both rigid rods and flexible chains [25,26]. The extension of our approach to explicitly treat such semiflexible polymers in the bulk and under probe microrheology situations remains an open problem. Thus, standard bulk rheology measurements on rigid rod entangled synthetic or biopolymer systems are urgently needed to definitively test our non-classical predictions. We also encourage new simulations be performed for needle fluids, spanning the pre-overshoot to steady flow regimes to test our inter-connected predictions of how orientation, tube diameter, and other quantities evolve with strain and Wi.

 	The theoretical frontier within the context of our force-level approach is now entangled flexible chain liquids at Rouse Wi numbers exceeding unity, i.e., flow rates where polymers strongly stretch. Here, Wang and coworkers [19-21] have proposed conceptually new ideas associated with an intermolecular ``grip force'' required for chains to stretch under deformation, and a force imbalance condition (``elastic yielding'') required before stretched chains can retract, corresponding to effective an entropic barrier to restore contour length (near) equilibrium. A fundamental theoretical basis for these ideas does not presently exist, nor is it obvious how (or if) they can be reconciled [15] with the phenomenological DE and GLaMM models. Progress on these fundamental issues is urgently needed. In the slow flow regime, recent Brownian dynamics simulations have found non-classical behavior [27] (but, importantly, another simulation did not [28]) challenging the DE idea of rapid Rouse-like retraction. In the fast flow high Rouse Wi regime, a remarkable fractional power-law scaling of the stress overshoot stress and strain has been observed [20, 27], which is in strong disagreement with existing phenomenological tube models. As will be reported in paper III of this series [30], the grip force concept has now been microscopically formulated, and a model for non-Rouse chain retraction constructed, which have been quantitatively confronted with experiments and simulations. Finally, by building on the advances in papers I, II and III, we aim to construct a microscopic, force-level theory for entangled chain liquids applicable at all values of Wi which includes a self-consistent treatment of the transverse and longitudinal aspects of entanglement physics. Work is in progress in this direction, and upon completion will be the subject of paper IV of this series.

\acknowledgements{KSS thanks Shi-Qing Wang for many stimulating, informative, motivating and spirited discussions (and arguments) over the years. Helpful discussions and correspondence with Dimitris Vlassopoulos and Peter Olmsted are also gratefully acknowledged. This work in its later stages was partially supported by DOE-BES via the Frederick Seitz Materials Research Laboratory (KSS) and the Advanced Materials Fellowship of the American Philosophical Society (DMS). }

\end{document}